\documentclass{rspublic}
\usepackage{amsmath}
\input{epsf}

\newcommand\bfu{{\bf u}}
\newcommand\bfb{{\bf b}}
\newcommand\bfz{{\bf z}}
\newcommand\bfxhat{\hat{\bf x}}
\newcommand\bfyhat{\hat{\bf y}}
\newcommand\bfzhat{\hat{\bf z}}
\newcommand\bfU{{\bf U}}
\newcommand\bfF{{\bf F}}

\newcommand\bfB{{\bf B}}
\newcommand\bfx{{\bf x}}

\newcommand\av[1]{\langle #1 \rangle}

\newcommand\half{\frac{1}{2}}
\newcommand\etal{\textit{et al}.\ }

\begin{document}

\title[Dynamo Action in an Imposed Field]{Dynamo Action in the Presence of an Imposed Magnetic Field}

\author[D.W. Hughes \& M.R.E. Proctor]{D.W. Hughes$^1$ \& M.R.E. Proctor$^2$}

\affiliation{$^1$Department of Applied Mathematics, University of Leeds, Leeds LS2 9JT, UK \\
$^2$D.A.M.T.P., Centre for Mathematical Sciences, University of Cambridge, Cambridge CB3 0WA, UK}

\label{firstpage}

\maketitle

\begin{abstract}{dynamo theory; fast dynamo action; magnetohydrodynamic instability}
We consider the linear stability to three-dimensional perturbations of two-dimensional nonlinear magnetohydrodynamic basic states obtained from a specified forcing function in the presence of an imposed initially uniform magnetic field of strength $B_0$. The forcing is chosen such that it drives the CP flow of Galloway \& Proctor (1992) when $B_0=0$. 
We first examine the properties of these basic states and their dependence on $B_0$ and on the magnetic Reynolds number $Rm$. The linear stability of these states is then investigated. It is found that at a given $Rm$ the presence of a background field is stabilising. The results also allow us to speculate that at a fixed value of $B_0$ the growth of the unstable perturbations is `fast', in the sense that the growth rate becomes independent of $Rm$ as $Rm \rightarrow \infty$.

\end{abstract}

\section{Introduction}\label{sec:intro}

The problem of the sustained generation of magnetic fields by fluid motion (dynamo action) is of immense importance in understanding the internal dynamics of the Sun, Earth and planets. Most research to date has focused on the derivation of conditions under which small initial fields, \textit{with no net flux}, can grow as a result of dynamo action (the kinematic dynamo problem), and on the size they reach before dynamical effects supervene and halt the growth of magnetic energy (the dynamical or magnetohydrodynamical dynamo problem). For these classical dynamo problems there is a clear distinction between, on the one hand, the non-magnetic case and, on the other, two equivalent dynamo states, related by having opposite signs of the magnetic field ${\bf B}$, but with the same velocity ${\bf u}$. 


It is easy, however, to conceive of situations in nature in which the ambient flux is unlikely to be zero, and in which small-scale fluid motions exist in a background of a large-scale field that, measured over the appropriate scales, has a net flux. On the Sun, for example, the supergranular and granular convective motions are threaded by a magnetic field of a much larger scale emerging from deeper in the solar interior (and, most likely, the result of another dynamo mechanism --- possibly here one of zero net flux). In a different astrophysical context, the significant magnetic fields of the jovian moons Io and Ganymede may result from their magnetohydrodynamic (MHD) processes taking place within the magnetic field of Jupiter (see Sarson \etal 1997).

It is therefore important to enquire into the analogue of the dynamo state when the imposed field is not zero. When the magnetic Reynolds number $Rm$ (defined by $Rm={\cal UL}/\eta$, where ${\cal U}$, ${\cal L}$ are typical velocity and length scales and $\eta$ is the magnetic diffusivity) is large, the fluctuating fields induced have strength much larger than the imposed field strength $B_0$, and indeed for sufficiently small $B_0$ the solutions are almost independent of $B_0$, apart from the imposed flux, which is small (relative to the rms field). Nonetheless, one might expect, on grounds of continuity, that there is more than one possible state of the system; one deriving continuously from the zero-field case, and two others (at least) from the two equilibrated dynamo solutions when $B_0=0$. This in turn leads to the general question: as $Rm$ is increased, with $B_0 \ne 0$, what is the nature of the transition that is analogous to the dynamo instability? Because of the pre-existence of a Lorentz force field, this question cannot be answered simply by looking at the dynamo properties of a velocity field modified by the Lorentz forces. Instead, the stability calculation involves, in an essential way, the equation of motion as well as the induction equation. This dichotomy has also recently been noted by Cattaneo \& Tobias (2008) and Tilgner \& Brandenburg (2008).

One well-studied system in which the distinction naturally arises between the amplification of a background field (of non-zero flux) and the dynamo generation of a field of zero flux is that of magnetohydrodynamic convection. It is well known that when $Rm$ is sufficiently small, the field structure depends vitally on the imposed field, and that all the field would disappear if the imposed flux were removed. This is the classical problem of \textit{magnetoconvection}, which has been studied over many years in a number of guises (see the reviews by Proctor \& Weiss (1982), Hughes \& Proctor (1988) and Proctor (2005), and references therein). However, if $Rm$ is sufficiently large then convectively driven flows can act as a dynamo. Early analytical work by Childress \& Soward (1972) and Soward (1974) on rapidly rotating convection confirmed the possibility of dynamo action in that case. Cattaneo (1999), following on from the pioneering study of Meneguzzi \& Pouquet (1989), showed numerically that vigorous convection (with $Rm$ of the order of a few hundred) could result in small-scale dynamo action (with no significant mean field component) even in the absence of rotation. Cattaneo \& Hughes (2006) carried out a detailed numerical investigation of the rotating case. In all these simulations a statistically steady state is reached, in which the magnetic energy is smaller than, but of the same general size as, the kinetic energy; the vigour of the convection is somewhat reduced but its Eulerian properties do not seem much changed.

Dynamo simulations are of course conducted without an imposed magnetic field. A detailed study of the effect of an imposed field on a convective state sufficiently vigorous to act as a dynamo has been performed by Cattaneo \etal (2003). This shows a transition from a convective dynamo to something indistinguishable from magnetoconvection as the strength of the imposed field is increased. Because the flows are disordered, and no obvious symmetries are broken as the fields and flow change, it is hard to detect any symmetry-breaking effect associated with the analogue of the onset of dynamo action when an imposed magnetic field is present. Nonetheless one would expect some such transition to be present as argued above. Even if the broad features of the fields do not change much when the transition occurs, it is important to acquire some understanding of the nature of the solution space, and how it depends on the parameters of the problem. However for the general case it is not clear how to proceed at present.

The idea of this paper is to sidestep these difficulties by investigating a simple system in which the basic state has a symmetry that will be broken by the instability. Thus the original and bifurcated solutions can be clearly distinguished. In our particular case we consider the evolution of three-dimensional linear perturbations to a two-dimensional magnetohydrodynamic basic state. Our principal goal is to see the effects of a background field on a well-studied and important instability. The new effect can be characterised as an extension of the classical kinematic dynamo problem, for which the basic state consists of a prescribed velocity field but no magnetic field. In order to build upon work performed on the classical problem, it makes sense to consider a basic state that, in the absence of an imposed field, reduces to a flow whose kinematic dynamo properties are well studied. Furthermore, since astrophysical interest is in the properties of dynamos at high $Rm$, we also wish to consider a case for which the basic flow field (i.e.\ that with $B_0 =0$) acts as a \textit{fast} dynamo --- namely, magnetic field growth rates become independent of the magnetic diffusivity as $Rm \rightarrow \infty$. This enables us to explore the analogue of `fastness' when the imposed field is non-zero. For these reasons we consider basic MHD states driven by a forcing such that, in the absence of an imposed field, the flow reduces to the so-called CP flow of Galloway \& Proctor (1992), whose dynamo properties are well understood (see also, for example, Cattaneo \etal 1995b; Hughes \etal 1996).

In \S\ref{sec:mf} we first discuss the mathematical formulation of the general extended dynamo problem and then look in particular at the modified Galloway-Proctor problem. In \S\ref{sec:bs} we discuss in detail the MHD states, first in terms of bounds on average measures of the field and flow, and then by considering numerical solutions of the basic state. Section~\ref{sec:da} contains the key results of the paper describing the linear stability of a range of basic states, differing in $Rm$ and in the strength of the background field. The results and their implications are summarised in \S\ref{sec:conc}.


\section{Mathematical Formulation}\label{sec:mf}

As explained above, in order to study unequivocally the possibility of linear dynamo action in the presence of a background magnetic field it is necessary that the basic state and the perturbations are qualitatively different, and hence can be unambiguously distinguished. The mathematical formulation of the general problem is as follows. 
As basic states we consider the long-time, stationary (but typically time-dependent) states resulting from a forcing ${\bf F}(\bfx,t)$ in the presence of an initially uniform magnetic field, where the forcing is chosen so as to impose a certain symmetry on the basic state. For simplicity we suppose the fluid to be incompressible. The evolution of the basic state flow $\bfU$ and basic state magnetic field $\bfB$ are then determined by the following dimensionless equations:
\begin{eqnarray}
\frac{\partial \bfU}{\partial t} + \bfU \cdot \nabla \bfU & = & - \nabla \Pi + \bfB \cdot \nabla \bfB + Re^{-1} \nabla^2 \bfU + \bfF , \label{eq:bs1}\\
\frac{\partial \bfB}{\partial t} + \bfU \cdot \nabla \bfB & = & \bfB \cdot \nabla \bfU + Rm^{-1} \nabla^2 \bfB , \label{eq:bs2}\\
\nabla \cdot \bfU &=& 0, \label{eq:bs3}\\
\nabla \cdot \bfB &=& 0, \label{eq:bs4}
\end{eqnarray}
where $\Pi$ denotes the total pressure (gas $+$ magnetic), and $Re$ and $Rm$ are the fluid and magnetic Reynolds numbers. It should be noted that the Lorentz force term in (\ref{eq:bs1}) has no pre-factor, implying that magnetic fields are measured in terms of the equivalent Alfv\'en velocity and not in terms of the imposed field strength $B_0$; this is the convention used in dynamo theory, though not in magnetoconvection.

Linear perturbations $\bfu$, $\bfb$ and $\pi$ to the (nonlinear) basic state described by $\bfU$, $\bfB$ and $\Pi$ are then governed by the following equations:
\begin{eqnarray}
\frac{\partial \bfu}{\partial t} + \bfU \cdot \nabla \bfu + \bfu \cdot \nabla \bfU & = & - \nabla \pi + \bfB \cdot \nabla \bfb + \bfb \cdot \nabla \bfB + Re^{-1} \nabla^2 \bfu , \label{eq:pert1} \\
\frac{\partial \bfb}{\partial t} + \bfU \cdot \nabla \bfb + \bfu \cdot \nabla \bfB & = & \bfB \cdot \nabla \bfu + \bfb \cdot \nabla \bfU + Rm^{-1} \nabla^2 \bfb , \label{eq:pert2} \\
\nabla \cdot \bfu &=& 0, \label{eq:pert3} \\
\nabla \cdot \bfb &=& 0. \label{eq:pert4} 
\end{eqnarray}
Kinematic dynamo action in this context therefore corresponds to an average exponential growth of both the perturbation magnetic energy \textit{and} the perturbation kinetic energy. Mathematically the problem involves the solution of the six equations (\ref{eq:bs1}), (\ref{eq:bs2}), (\ref{eq:bs3}), (\ref{eq:pert1}), (\ref{eq:pert2}) and (\ref{eq:pert3}), subject to the solenoidal constraints (\ref{eq:bs4}) and (\ref{eq:pert4}); this is in contrast to the classical kinematic dynamo problem, for which $\bfU$ is prescribed, $\bfB = \bfu =0$, and only (\ref{eq:pert2}) needs to be solved, subject to the constraint (\ref{eq:pert4}). Because the coupled system (\ref{eq:pert1}) -- (\ref{eq:pert3}) is so different from the induction equation alone it is difficult to predict \textit{a priori} the stability properties of the system when $B_0$ is not small.

For our study we adopt a two-dimensional spatially-periodic forcing $\bfF (x,y,t)$ and impose that the basic state is similarly $z$-independent and with the same spatial periodicity as $\bfF$ in the $xy$-plane. We may then seek three-dimensional perturbations of the form
\begin{equation}
\bfu (x,y,z,t) = {\hat \bfu} (x,y,t) \exp(ikz)
\label{eq:pert_sol}
\end{equation}
(and similarly for $\bfb$), where $k$ is a real wavenumber. It should be noted here that the fields $\bfb$ and $-\bfb$ are mapped into each other through a shift in $z$, and we are therefore unable to discuss the interesting secondary question, generally applicable to problems of this type, as to whether the sign-change symmetry of the instability is broken by the presence of a background field.

For our specific choice of $\bfF$ we adopt the forcing that, in the absence of a background field and subject to stability considerations, drives the CP flow of Galloway \& Proctor (1992). We represent the velocity in poloidal and toroidal parts by writing
\begin{equation}
\bfU = \nabla \times (\psi{\bf \hat z}) + w {\bf \hat z}\equiv \bfU_H + w {\bf \hat z},
\label{eq:GP1}
\end{equation}
where, here and below, we use the subscript $H$ to refer to the `horizontal' components (i.e.\ those perpendicular to $\bfzhat$). For the basic CP flow we have
\begin{equation}
\psi=w = \sqrt{3/2} \left( \cos (x+ \cos t) + \sin (y + \sin t) \right) .
\label{eq:GP2}
\end{equation}
The flow is maximally helical (Beltrami), i.e.\ $\bfU$ is parallel to $\nabla \times \bfU$; as such, it can be driven by the forcing $\bfF$ given by
\begin{equation}
\bfF \equiv \bfF_H + F {\bf \hat z}=(\partial_t - Re^{-1} \nabla^2 ) \bfU .
\label{eq:force}
\end{equation}
(The pressure gradient does not appear because of the Beltrami property of $\bfU$.) In contrast to the steady three-dimensional ABC flow analysed by Podvigina \& Pouquet (1994), which becomes unstable for $Re \gtrsim 13$ (for $A=B=C=1$), the forcing (\ref{eq:force}) restricted to two-dimensional flows certainly yields the flow (\ref{eq:GP1}, 
\ref{eq:GP2}) for $Re$ of $O(10^3)$ --- and indeed maybe does so for much higher $Re$. However, the time scale for the decay of transient behaviour scales with $Re$, and thus transients are extremely long-lived at high $Re$.

The Galloway-Proctor flow (\ref{eq:GP2}) is a time-dependent extension of the steady, maximally helical cellular flow first considered by Roberts (1970). It has sizeable regions of Lagrangian chaos, as measured by its Lyapunov exponents (see, for example, Cattaneo \etal 1995a), and, from the numerical evidence, appears to act as a fast dynamo (Galloway \& Proctor 1992). For sufficiently high $Rm$ ($\gtrsim 100$) the dynamo growth rate is maximised for an $O(1)$ value of the wavenumber $k$ ($k = 0.57$) and takes a value close to $0.3$.

\section{The Basic State}\label{sec:bs}

In this section we consider the (time-dependent) basic states that emerge as the long-time solutions to equations 
(\ref{eq:bs1}) -- (\ref{eq:bs4}), with initial conditions of $\bfU =0$ and $\bfB = B_0 \hat \bfx$. These are described by three parameters; the strength of the imposed field $B_0$, the fluid Reynolds number $Re$ and the magnetic Reynolds number $Rm$.

From (\ref{eq:force}) it can be seen that in order to drive the same flow at different values of $Re$ the force clearly has to be $Re$-dependent. Comparisons between runs with different values of $Re$ are therefore somewhat problematic since they are not simply related by a change in viscosity. Consequently we choose to keep $Re$ fixed for all of our calculations and to vary $Rm$ and $B_0$. In order to keep the duration of any transient phase in the evolution of the flow reasonably short, we opt for the value of $Re=1$.

\subsection{Bounds}
\label{subsec:bounds}

These basic states are interesting in their own right; and one can in fact derive rigorous limits on the averages of the fluctuating fields and flows. Define 
\begin{equation}
\av{{ \bf\cdot}} = \frac{1}{4 \pi^2 T} \int_0^T \int_0^{2 \pi} \int_0^{2 \pi} {{\bf\cdot}} \, dx dy dt,
\label{eq:ave}
\end{equation}
where $T$ is the temporal period of the solution (and where, if the solution is not periodic, we take $T\rightarrow\infty$). We write $\bfB=\bfB_0+\bfB_H+ B{\bf \hat z}$, $\bfU=\bfU_H+ w{\bf \hat z}$ and first consider the $z$-components of (\ref{eq:bs1}) and (\ref{eq:bs2}). If we multiply these equations by $w$ and $B$ respectively and perform the spatial and temporal averages, we obtain
\begin{align}
\bfB_0\cdot\av{ w\nabla B}+\av{ w\bfB_H\cdot\nabla B}&-\frac{1}{Re}\av{|\nabla w|^2}+\av{ w F}=0, \label{eq:av1}\\
\bfB_0\cdot\av{ w\nabla B}+\av{ w\bfB_H\cdot\nabla B}&+\frac{1}{Rm}\av{|\nabla B|^2}=0 . \label{eq:av2}
\end{align}

Integrating by parts and using the spatial periodicity gives $\av{w\nabla B}=-\av{B\nabla w}$, $\av{w\bfB_H\cdot\nabla B}=-\av{B\bfB_H\cdot\nabla w}$. We may then use the Schwarz inequality twice in (\ref{eq:av2}) to derive the inequality
\begin{equation} 
\frac{1}{Rm} \av{|\nabla B|^2} 
\leq \av{|\nabla w|^2}^\half \left[ B_0 \av{ B^2}^\half 
+ \av{|\bfB_H|^4}^\frac{1}{4} \av{B^4}^\frac{1}{4} \right] , \label{eq:bound1}
\end{equation}
which, in principle, gives a lower bound on the viscous dissipation of the $z$-component of the flow. 
Another result is found by subtracting (\ref{eq:av1}) from (\ref{eq:av2}), to give
\begin{equation}
\frac{1}{Re}\av{w^2}+\frac{1}{Rm}\av{B^2}\leq \frac{1}{Re}\av{|\nabla w|^2}+\frac{1}{Rm}\av{|\nabla B|^2} \leq \av{w^2}^\frac{1}{2} {\cal F}\leq \av{|\nabla w|^2}^\frac{1}{2}{\cal F},
\label{eq:bound2}
\end{equation}
where ${\cal F}^2=\av{F^2}$, and where we have used the fact, here and elsewhere, that for $2\pi\times 2\pi$ periodic functions $g(x,y)$ of zero mean, $\av {|\nabla g|^2} \geq \av{g^2}$.

If we consider the horizontal components then we can derive results that are directly analogous to (\ref{eq:bound1}) and (\ref{eq:bound2}), namely
\begin{equation} 
\frac{1}{Rm} \av{|\nabla \bfB_H|^2} 
\leq \av{|\nabla \bfU_H|^2}^\half \left[ B_0 \av{ |\bfB_H|^2}^\half 
+ \av{|\bfB_H|^4}^\frac{1}{2} \right]\label{eq:bound6} 
\end{equation}
and
\begin{equation}
\begin{split}
\frac{1}{Re}\av{|\bfU_H|^2}+\frac{1}{Rm}\av{|\bfB_H|^2} &\leq\frac{1}{Re}\av{|\nabla\bfU_H|^2}+\frac{1}{Rm}\av{|\nabla\bfB_H|^2} \\ &\leq \av{|\bfU_H|^2}^\frac{1}{2}{\cal F}_H \leq \av{|\nabla \bfU_H|^2}^\frac{1}{2}{\cal F}_H,
\label{eq:bound7}
\end{split}
\end{equation}
where ${\cal F}_H=\av{|\bfF_H|^2}^\frac{1}{2}$. (For the forcing (\ref{eq:force}), ${\cal F}_H$ = ${\cal F}$ = $\left( 3 + 3/(2Re^2) \right)^{1/2}$.) We can also use (\ref{eq:bound2}) and (\ref{eq:bound7}) to show that 
\begin{equation}
\av{B^2}\leq\av{|\nabla B|^2}\leq \frac{1}{4}Re Rm{\cal F}^2
\label{eq:bound8}
\end{equation}
and
\begin{equation}
\av{|\bfB_H|^2}\leq\av{|\nabla\bfB_H|^2}\leq \frac{1}{4}Re Rm{\cal F}_H^2 .
\label{eq:bound9}
\end{equation}

We can make further progress with the horizontal components of the induction equation. Because of the two-dimensionality we can write $\bfB_H=\nabla \times (A\hat{\bfz})$, where $A$ satisfies
\begin{equation}
\frac{\partial A}{\partial t}+\bfU_H\cdot\nabla A= \hat{\bfz}\cdot(\bfU_H\times\bfB_0)+Rm^{-1}\nabla^2 A .
\label{eq:bound3}
\end{equation}
After multiplication of this equation by $A$ and averaging, we obtain the result
\begin{equation}
Rm^{-1}\av{|\bfB_H|^2}\leq B_0 \av{|\bfU_H|^2}^\half\av{A^2}^\half\leq B_0 \av{|\bfU_H|^2}^\half\av{|\bfB_H|^2}^\half ,
\label{eq:bound4}
\end{equation}
so that
\begin{equation}
Rm^{-1}\av{|\bfB_H|^2}^\half \leq B_0\av{|\bfU_H|^2}^\half.
\label{eq:bound5}
\end{equation}
We shall investigate the sharpness of these results in \S\ref{sec:bs}$\ref{subsec:numres}$ below.

\subsection{Numerical results}
\label{subsec:numres}

In this subsection we describe the results of numerical simulations of the basic state; $Re=1$ in every case. We look in detail at two values of the magnetic Reynolds number, $Rm =100$ and $Rm = 1000$, and at a range of values of the imposed field strength $B_0$. The equations (\ref{eq:bs1}) -- (\ref{eq:bs4}) are integrated in time from an initially static state until a time-periodic or statistically steady state is reached. In the absence of a background field the kinetic energy takes the value $1.5$ (independent of time); the energy of the imposed magnetic field, $\langle B_0^2 \rangle /2$, should thus be compared with this value. (It should be noted that in some of the runs there is a very fast initial stage where the magnetic energy rises rapidly at the start of the calculation.)

\subsubsection{$Rm = 100$}

\begin{figure}
\vskip 0cm
\hskip 0cm
\epsfxsize 12.5cm
{\centerline{\epsffile{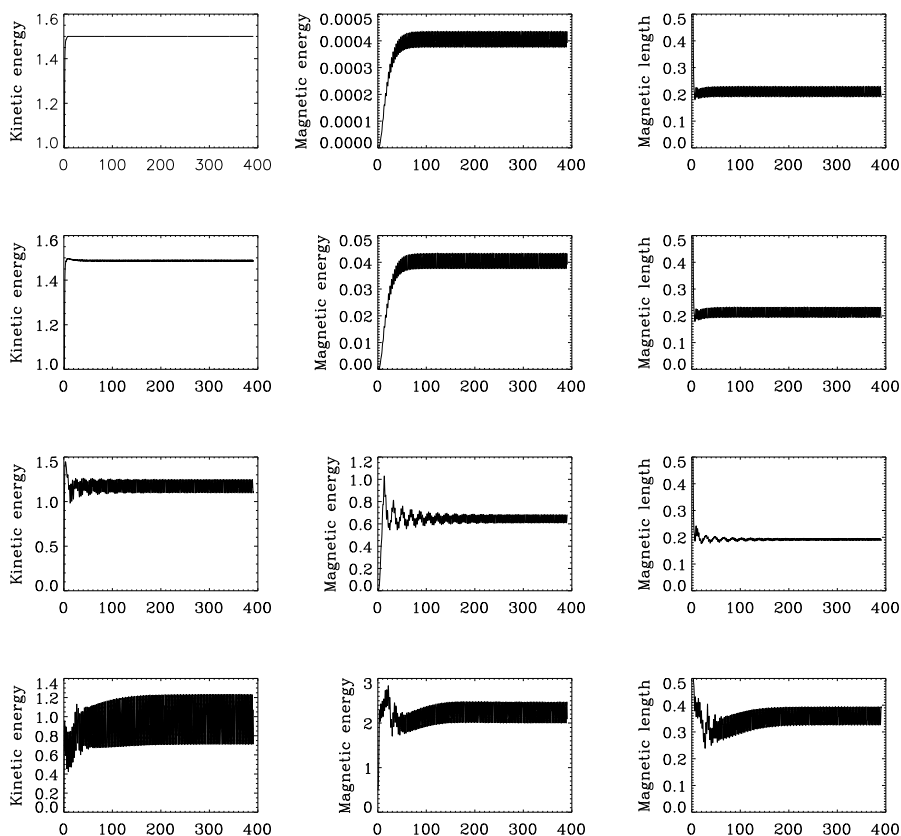}}}
\caption{Plots of the kinetic energy, magnetic energy and magnetic length $\ell_B$ versus time for the basic states with $Rm=100$ and (from top to bottom) imposed fields $B_0=0.001$, $0.01$, $0.1$ and $1$.
} 
\label{fig:energies_100}
\end{figure}

\begin{figure}
\vskip 0cm
\hskip 0.1cm
\epsfxsize 12cm
{\centerline{\epsffile{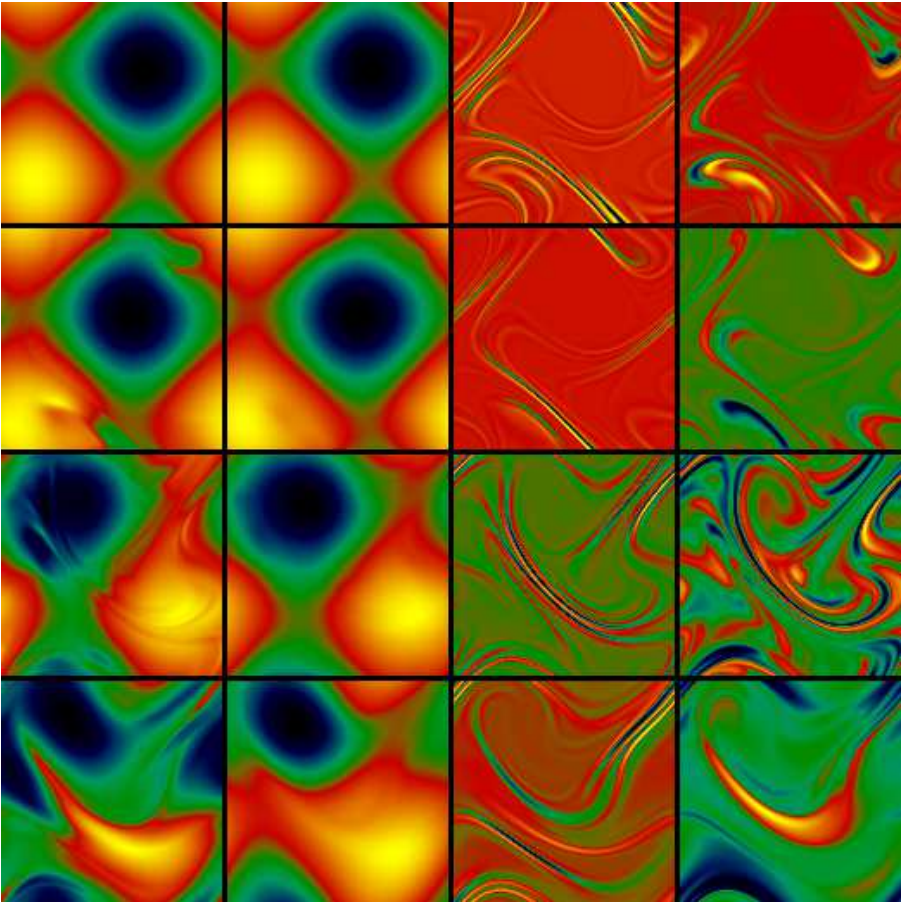}}}
\caption{Snapshots (all scaled individually) of (from left to right) the $z$-components of the vorticity, velocity, current and magnetic field. $Rm = 100$ and the imposed field strengths are (from top to bottom) $B_0 =0.001$, $0.01$, $0.1$ and $1$.
} 
\label{fig:basic_state}
\end{figure}

In figure~\ref{fig:energies_100} are plotted, for $Rm = 100$ and $B_0=0.001$, $0.01$, $0.1$ and $1$, the kinetic energy, the total magnetic energy (including that of the imposed field $B_0$) and the magnetic length $\ell_B$, where $\ell_B^2\equiv\av{|\bfB-\bfB_0|^2}/\av{|\nabla \times \bfB |^2}$. The last gives information on the typical dissipative scale of the magnetic field. For the weakest field strength considered, the kinetic energy remains steady and the magnetic energy is periodic, with twice the frequency of the forcing. At the larger field strengths, the kinetic energy is periodic and the magnetic energy either periodic or quasi-periodic.

Clearly for $B_0 = 0.001$ the field remains kinematic throughout the evolution. Although the magnetic energy has been considerably amplified (by a factor dependent on $Rm$), it remains very small in comparison with the kinetic energy. The amplitude of the field in the saturated state is determined by diffusion in this regime, the Lorentz forces being negligible; a series of purely kinematic runs reveals that in the absence of the Lorentz force the magnetic energy scales as $Rm^\gamma$ with $\gamma \approx 1.4$. For $B_0 = 0.001$, $\ell_B$ is also determined solely by diffusion, with $\ell_B \propto Rm^{1/2}$ in the kinematic regime. For $B_0=0.01$ the field is just dynamically active, with small oscillations visible in the kinetic energy. When $B_0 =0.1$, however, the kinetic energy is noticeably reduced and the field is obviously dynamical in its behaviour --- the small-scale fields are amplified to a dynamically significant strength despite the fact that the energy in the imposed field is still very weak in comparison with the kinetic energy of the flow. For $B_0=1$ even the imposed field has equipartition strength and no aspect of the evolution can be thought of as kinematic. It should be noted that for our choice of forcing, i.e.\ that given by (\ref{eq:GP1}) -- (\ref{eq:force}), a unidirectional imposed magnetic field can never suppress the flow completely. In the limit of a very strong field (in the $x$-direction) $\bfU \cdot \bfyhat$ is suppressed, as is the $x$-dependent component of $\bfU \cdot \bfzhat$; the magnetic energy of the final state is essentially $B_0^2/2$ and the kinetic energy takes the value $0.75$.

Figure~\ref{fig:basic_state} shows snapshots of the structure of the flow and field in the basic state for the same imposed fields as in figure~\ref{fig:energies_100}. For $B_0=0.001$ the field is so weak that the flow is simply the Galloway-Proctor flow given by (\ref{eq:GP1}, \ref{eq:GP2}); the evolution of the field is kinematic and the extent of the thin current sheets is determined solely by diffusion. Small deviations from the Galloway-Proctor flow start to appear when $B_0=0.01$. These become more pronounced for higher $B_0$, with an accompanying thickening of the magnetic structures.

\begin{figure}
\vskip 0cm
\hskip 0cm
\epsfxsize 10cm
{\centerline{\epsffile{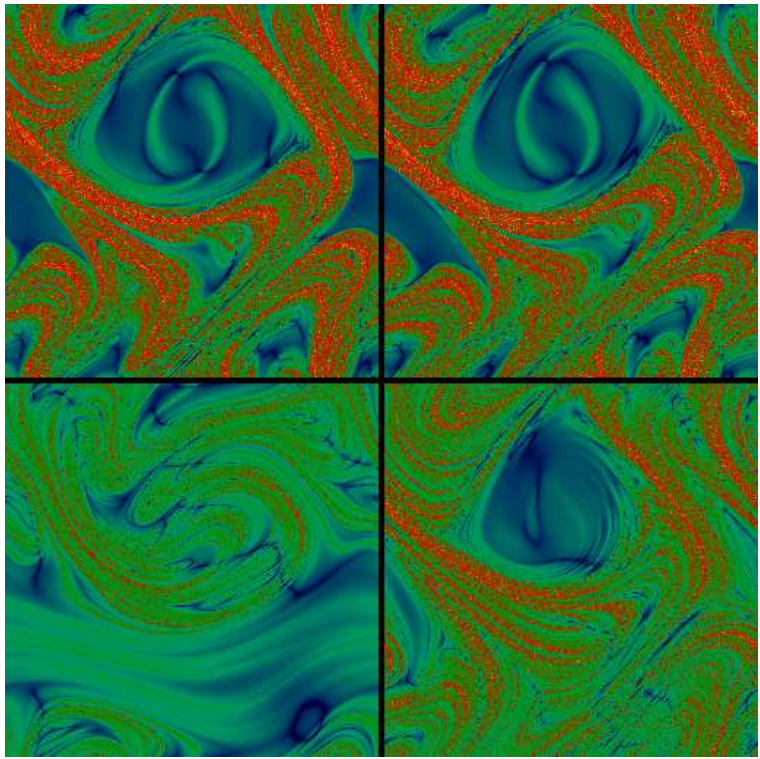}}}
\caption{Plots of the finite-time ($t = 25$) Lyapunov exponents for $Rm = 100$ and imposed fields (clockwise from top left) $B_0=0$, $B_0=0.01$, $B_0=0.1$, $B_0=1$.}
\label{fig:lyap_100}
\end{figure}

Given the importance for fast dynamo action (in the absence of $B_0$) of the chaotic nature of the driving flow, it is important to examine any changes in the chaotic properties of the flow brought about as $B_0$ is increased. For flows of the form $\bfu (x,y,t)$, exponential separation of neighbouring fluid elements takes place only in the $xy$-plane. Figure~\ref{fig:lyap_100} shows the finite-time Lyapunov exponents for four values of the imposed field. As expected, the chaotic nature of the fluid trajectories is suppressed as the imposed field strength is increased, but this is a fairly modest effect for values of $B_0$ up to $B_0=0.1$. The average value of the Lyapunov exponent falls from $0.162$ when there is no imposed field, to $0.157$, $0.151$ and $0.104$ when $B_0 = 0.01$, $0.1$ and $1$ respectively.

\begin{table}
\begin{tabular}{|c|c|c|c|c|}
\hline
$B_0$ & $0.001$ & $0.01$ & $0.1$ & $1.0$ \\ 
\hline
$\langle \bfU_H^2 \rangle $ & 1.5 & 1.488 & 1.152 & 0.865 \\
$\langle w^2 \rangle $ & 1.5 & 1.486 & 1.195 & 1.085 \\
$\langle | \nabla \bfU_H |^2 \rangle $ & 1.5 & 1.488 & 1.166 & 1.013 \\
$\langle | \nabla w |^2 \rangle $ & 1.5 & 1.486 & 1.23 & 1.232 \\
$\langle \bfB_H^2 \rangle $ & $2.079 \times 10^{-4}$ & 0.0203  & 0.512 & 1.75 \\
$\langle B^2 \rangle $ & $6.003 \times 10^{-4}$ & 0.0603  & 0.767  & 1.785  \\
$\langle | \nabla \bfB_H |^2 \rangle $ & 0.00795 & 0.767 & 19.862 & 15.99  \\
$\langle | \nabla B |^2 \rangle $ & 0.0109 & 1.066 & 14.86 & 12.24 \\
$\langle \bfB_H^4 \rangle $ & $1.55 \times 10^{-7}$ & 0.00151 & 0.926 & 4.34 \\
$\langle B^4 \rangle $ & $2.882 \times 10^{-6}$ & 0.0306 & 2.844 & 11.06 \\
\hline
${\cal R}(\ref{eq:bound1})  $ & 0.106 & 0.103 & 0.0984 & 0.0278  \\
${\cal R}(\ref{eq:bound2}ac)$ & 0.577 & 0.575 & 0.519  & 0.499   \\
${\cal R}(\ref{eq:bound2}bd)$ & 0.577 & 0.579 & 0.586  & 0.575   \\
${\cal R}(\ref{eq:bound2}bc)$ & 0.577 & 0.579 & 0.595  & 0.613   \\
${\cal R}(\ref{eq:bound6})  $ & 0.159 & 0.156 & 0.178  & 0.0466  \\
${\cal R}(\ref{eq:bound7}ac)$ & 0.577 & 0.575 & 0.508  & 0.447   \\
${\cal R}(\ref{eq:bound7}bd)$ & 0.577 & 0.578 & 0.596  & 0.549   \\
${\cal R}(\ref{eq:bound7}bc)$ & 0.577 & 0.578 & 0.599  & 0.595   \\
${\cal R}(\ref{eq:bound8}bc)$ & $9.69 \times 10^{-5}$ & 0.00948 & 0.132 & 0.109  \\
${\cal R}(\ref{eq:bound9}bc)$ & $7.07 \times 10^{-5}$ & 0.00682 & 0.177 & 0.142  \\
${\cal R}(\ref{eq:bound5})  $ & 0.118 & 0.117 & 0.0667 & 0.0142 \\
\hline
\end{tabular}
\label{table:bounds_100}
\caption{Average quantities required for the calculations of the bounds for $Rm = 100$, together with a measure of the tightness of the various bounds; ${\cal R}(\ref{eq:bound1})$, for example, denotes the ratio of the left hand side of inequality $(\ref{eq:bound1})$ to the right hand side; ${\cal R}(\ref{eq:bound2}ac)$ compares the first and third components of the multiple inequality $(\ref{eq:bound2})$. }
\end{table}

It is of interest to investigate the degree to which the divers bounds obtained in \S\ref{sec:bs}$\ref{subsec:bounds}$ are satisfied. Table~1 lists the average quantities involved, together with the ratio of the two sides in the various inequalities. It can thus be seen that the bounds described by the inequalities in (\ref{eq:bound2}) and (\ref{eq:bound7}) are reasonably tight (attained to better than within a factor of two) and also that they are fairly insensitive to the value of $B_0$. The bounds (\ref{eq:bound1}), (\ref{eq:bound6}), (\ref{eq:bound8}), (\ref{eq:bound9}) and (\ref{eq:bound5}) depend more strongly on $B_0$, but are shown to be rather loose for all the values of $B_0$ considered.

\subsubsection{$Rm = 1000$}

\begin{figure}
\vskip -0.8cm
\hskip 0cm
\epsfxsize 12cm
{\centerline{\epsffile{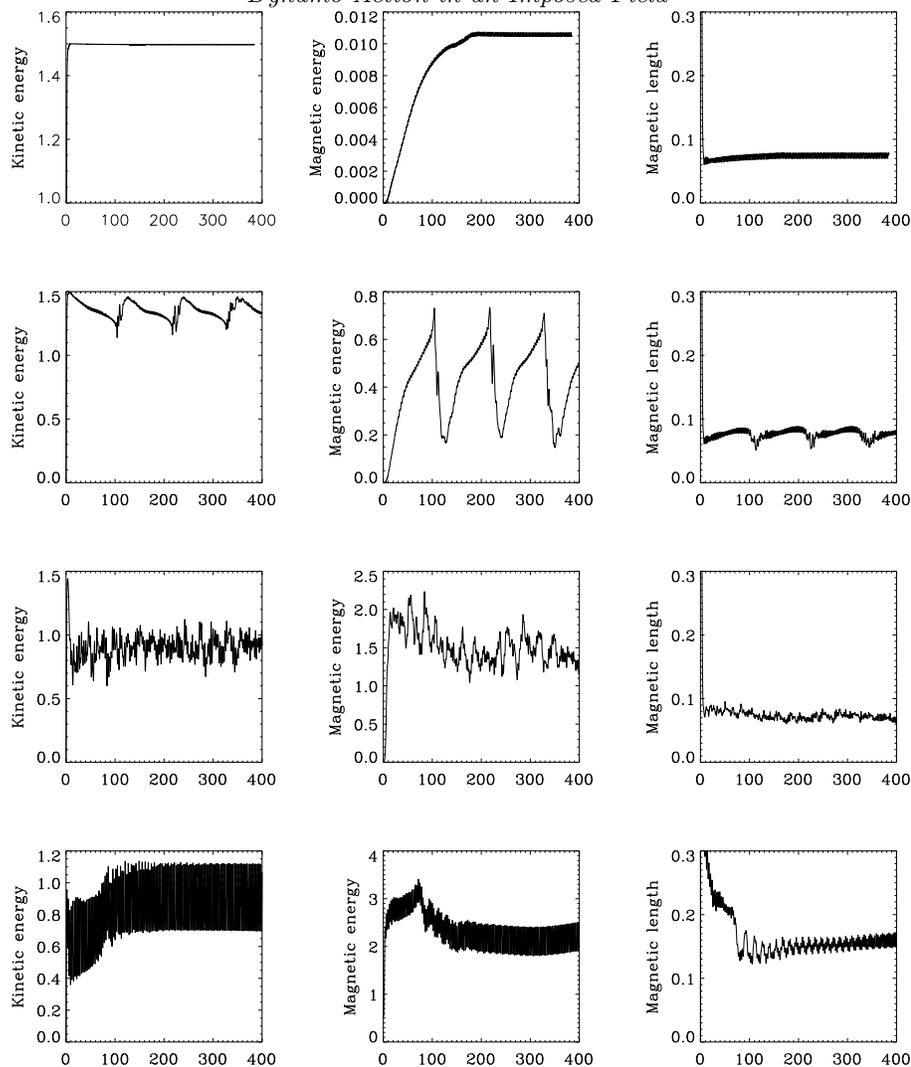}}}
\caption{Plots of the kinetic energy, magnetic energy and magnetic length $\ell_B$ versus time for the basic states with $Rm=1000$ and (from top to bottom) imposed fields $B_0=0.001$, $0.01$, $0.1$ and $1$.
}
\label{fig:energies_1000}
\end{figure}

When $Rm$ is increased to $Rm = 1000$ the basic state exhibits much greater temporal complexity.
Figure~\ref{fig:energies_1000} shows the evolution of the kinetic energy, magnetic energy and magnetic length for four different imposed fields. For $B_0 = 0.001$ the magnetic field is always kinematic, and the final state is simply periodic. Once the imposed field has strength $B_0 = 0.01$ the field is clearly dynamic; the kinetic and magnetic energies vary on both the short time scale of the forcing and on a much longer scale (of period 
$\approx 100$). Over the long period the magnetic energy grows gradually before plummeting drastically (and non-monotonically); the kinetic energy essentially does the opposite, declining slowly before rising abruptly. When $B_0 = 0.1$ the final state has a chaotic time dependence and is one of equipartition between the kinetic and magnetic energies. For the case of $B_0=1$, for which the imposed field is close to equipartition strength, the evolution shows a long-time transient before settling down to a quasi-periodic solution, which is manifested most clearly in $\ell_B$. The Lyapunov exponents are again reduced as the imposed field strength is increased; the average value of the finite-time Lyapunov exponent evaluated at $t=25$ falls from $0.168$ when $B_0=0.001$ (marginally, but probably not significantly, above its kinematic value) to $0.151$ when $B_0=0.01$, $0.107$ when $B_0=0.1$ and $0.073$ when $B=1$.

\begin{table}
\begin{tabular}{|c|c|c|c|c|}
\hline
$B_0$ & $0.001$ & $0.01$ & $0.1$ & $1.0$ \\ 
\hline
$\langle \bfU_H^2 \rangle $ & 1.498 & 1.368 & 0.877 & 0.763 \\
$\langle w^2 \rangle $ & 1.497 & 1.323 & 0.948 & 0.993 \\
$\langle | \nabla \bfU_H |^2 \rangle $ & 1.498 & 1.377 & 0.923 & 0.960 \\
$\langle | \nabla w |^2 \rangle $ & 1.497 & 1.361 & 1.024 & 1.209  \\
$\langle \bfB_H^2 \rangle $ & 0.00404 & 0.228 & 1.187 & 2.004 \\
$\langle B^2 \rangle $ & 0.01707 & 0.663 & 1.639 & 2.086 \\
$\langle | \nabla \bfB_H |^2 \rangle $ & 1.455 & 78.86 & 318.4 & 96.67 \\
$\langle | \nabla B |^2 \rangle $ & 2.279 & 72.48 & 241.6 & 59.8 \\
$\langle \bfB_H^4 \rangle $ & $8.505 \times 10^{-5}$ & 0.535 & 5.58 & 6.366 \\
$\langle B^4 \rangle $ & $4.463 \times 10^{-3}$ & 7.91 & 10.66 & 21.18 \\
\hline
${\cal R}(\ref{eq:bound1})  $ & 0.0746 & 0.0431 & 0.0822 & 0.0112  \\
${\cal R}(\ref{eq:bound2}ac)$ & 0.577  & 0.542  & 0.460  & 0.471    \\
${\cal R}(\ref{eq:bound2}bd)$ & 0.578  & 0.579  & 0.590  & 0.544    \\
${\cal R}(\ref{eq:bound2}bc)$ & 0.578  & 0.587  & 0.613  & 0.600    \\
${\cal R}(\ref{eq:bound6})  $ & 0.128  & 0.0913 & 0.134  & 0.025   \\
${\cal R}(\ref{eq:bound7}ac)$ & 0.577  & 0.551  & 0.442  & 0.413    \\
${\cal R}(\ref{eq:bound7}bd)$ & 0.578  & 0.585  & 0.609  & 0.508    \\
${\cal R}(\ref{eq:bound7}bc)$ & 0.578  & 0.587  & 0.625  & 0.570    \\
${\cal R}(\ref{eq:bound8}bc)$ & 0.00203 & 0.0644 & 0.215 & 0.0532   \\
${\cal R}(\ref{eq:bound9}bc)$ & 0.00129 & 0.0701 & 0.283 & 0.0859   \\
${\cal R}(\ref{eq:bound5})  $ & 0.0519  & 0.0408  & 0.0116 & 0.00162    \\
\hline
\end{tabular}
\label{table:bounds_1000}
\caption{As table~1 but for $Rm = 1000$. Averages for those cases for which the temporal evolution is chaotic are taken over an interval of length approximately $600$.}
\end{table}

Table~2 contains the appropriate information for the various bounds for $Rm = 1000$. The overall picture is the same as for the case of $Rm =100$. Bounds (\ref{eq:bound2}) and (\ref{eq:bound7}) are reasonably tight and fairly insensitive to $B_0$; the others are again more dependent on $B_0$, but are always rather loose.

\section{Dynamo Action}\label{sec:da}

In this section we examine the evolution of linear perturbations, of both $\bfb$ and $\bfu$, to eight basic states 
($Rm=100$, $1000$; $B_0= 0.001$, $0.01$, $0.1$, $1$), exploring a range of wavenumbers $k$. Once a stationary basic state has been attained, we evolve the perturbation equations (\ref{eq:pert1}) -- (\ref{eq:pert4}) in concert with the basic state equations (\ref{eq:bs1}) -- (\ref{eq:bs4}), and then use the long-time behaviour of the perturbations to obtain their average exponential growth (or decay). This is a well-defined procedure provided that averages are taken over times long compared with the time scale of variation of the basic state. Some care is therefore needed for the case of $Rm=1000$, $B_0=0.01$, for which the basic state varies on a very long timescale; this is addressed in more detail below.

\begin{figure}
\vskip -0.5cm
\hskip 0cm
\epsfxsize 11cm
{\centerline{\epsffile{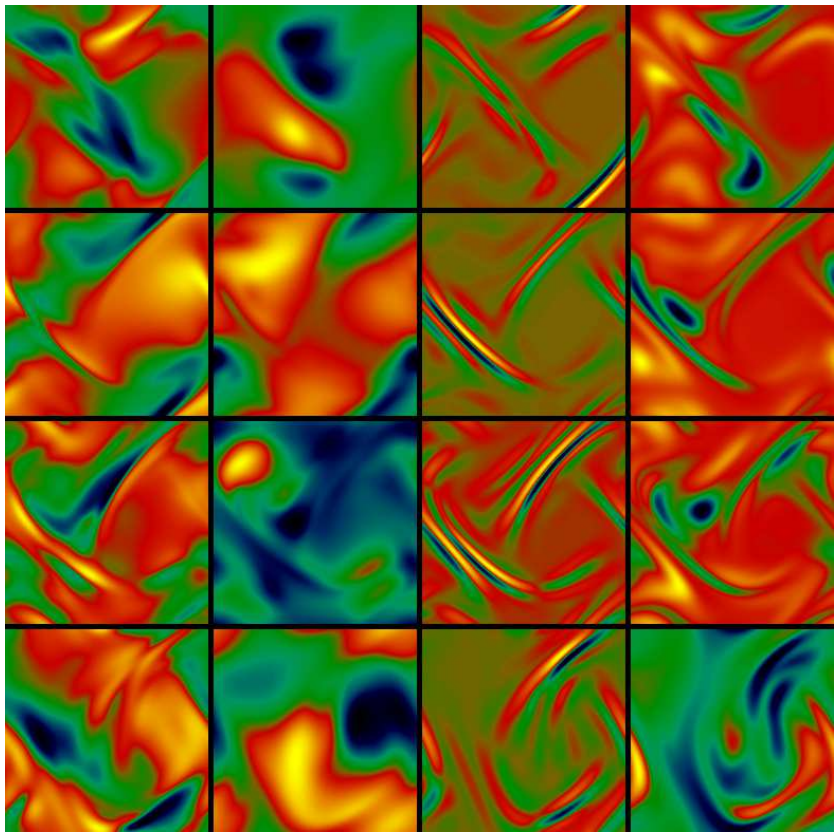}}}
\caption{Snapshots of the perturbation eigenfunctions for $Rm = 100$ at $z=0$ showing (from left to right) the 
$z$-components of the vorticity, velocity field, current and magnetic field, for the modes of maximum growth rates for (from top to bottom) $B_0 =0.001$, $0.01$, $0.1$ and $1.0$. Each image is scaled individually.
} 
\label{fig:eigfns_Rm100}
\end{figure}

\begin{figure}
\vskip -0.5cm
\hskip 0cm
\epsfxsize 11cm
{\centerline{\epsffile{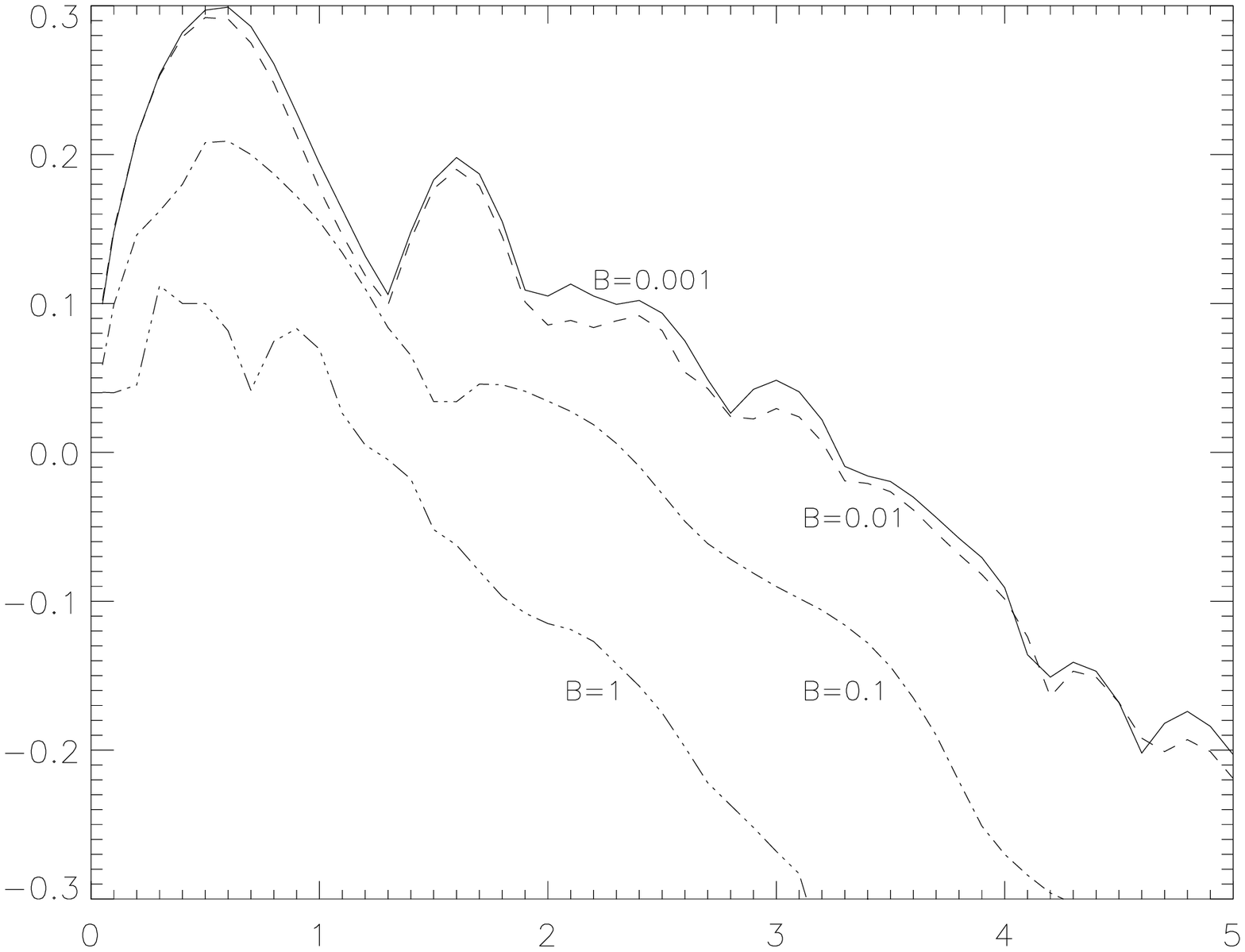}}}
\caption{Plots of the growth rates versus $k$ for $Rm = 100$ and for four different values of $B_0$ as shown.
} 
\label{fig:grates_100}
\end{figure}

\begin{table}
\begin{tabular}{ccccc}
\hline
\qquad  & \multicolumn{1}{c}{$B_0=0.001$} & \multicolumn{1}{c}{$B_0=0.01$} & \multicolumn{1}{c}{$B_0=0.1$} & \multicolumn{1}{c}{$B_0=1.0$} \\
\hline
$Rm=100$ & 37000 & 360 & 100 & 25 \\
$Rm = 1000$ & 7500 & 150 & 360 & 170 \\
\hline
\end{tabular}
\label{table:energy_ratios}
\caption{Ratio of magnetic to kinetic energies in the eigenmodes of maximum growth rate.}
\end{table}

In the absence of an imposed field, equations (\ref{eq:pert1}) and (\ref{eq:pert2}) decouple. For $Re=1$ and $B_0=0$ the flow is stable to three-dimensional disturbances and so perturbations in velocity decay exponentially; for wavenumbers $k$ such that the Galloway-Proctor flow acts as a dynamo, perturbations to the magnetic field grow exponentially. For all values of the imposed field strength considered, even when $B_0=1$ and the imposed field is essentially of equipartition strength, the magnetic energy of the perturbations greatly exceeds the kinetic energy. This can be seen in table~3, which contains the ratio of the energies for the modes of maximum growth rate. This indicates that the growth of disturbances is analogous to a dynamo instability rather than any possible hydrodynamic instability. For $Rm=100$ this ratio of energies is a monotonically decreasing function of $B_0$. Figure~\ref{fig:eigfns_Rm100} shows snapshots, for four values of $B_0$, of the eigenfunctions of the modes of maximum growth rate. The velocity and vorticity fields are large-scale, reflecting the low fluid Reynolds number. The thickness of the dominant magnetic structures increases with $B_0$ as the coupling between the flow and field becomes more pronounced.

\begin{figure}
\vskip -0.5cm
\hskip 0cm
\epsfxsize 11cm
{\centerline{\epsffile{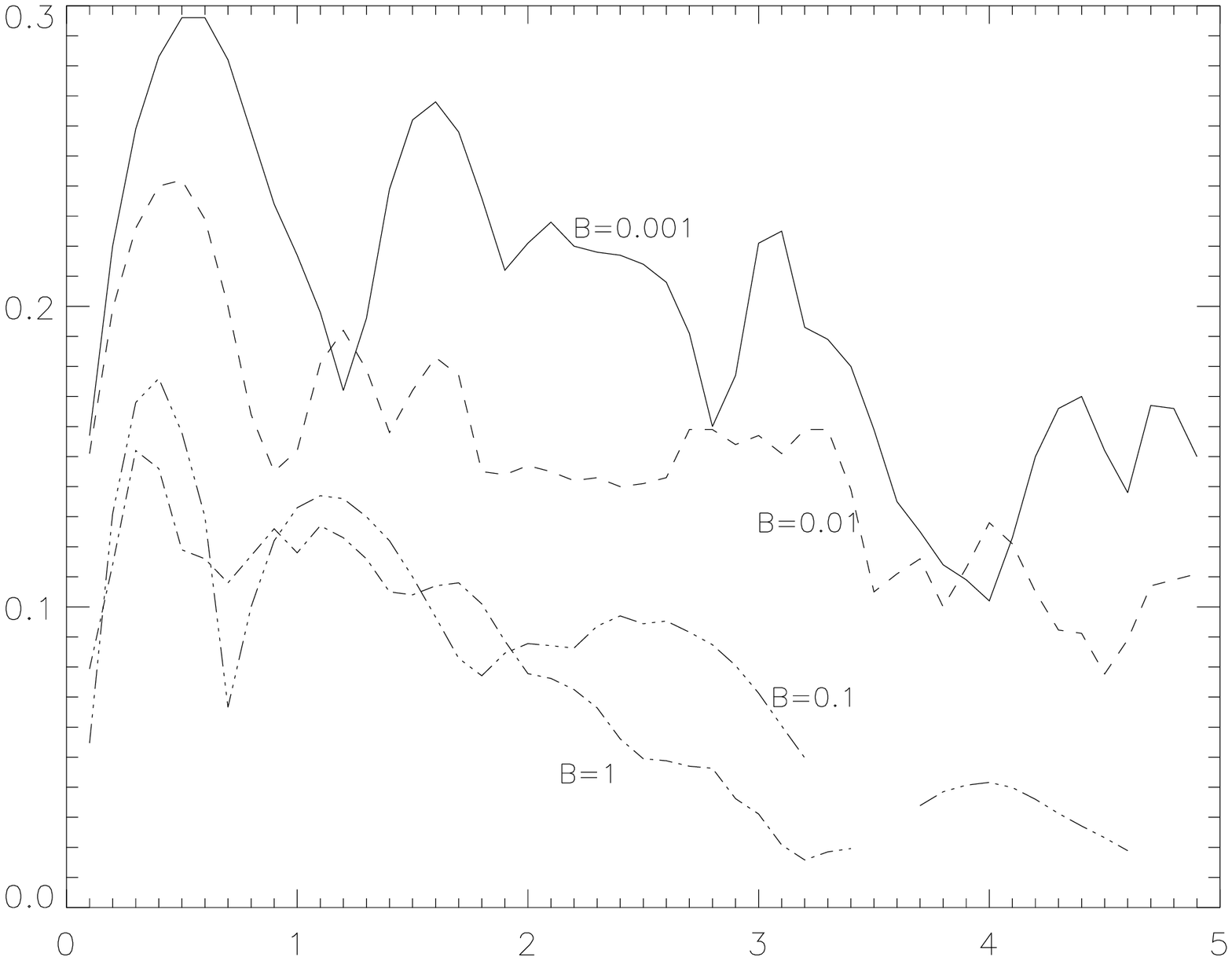}}}
\caption{Plots of the growth rates versus $k$ for $Rm = 1000$ and for four different values of $B_0$ as shown.
} 
\label{fig:grates_1000}
\end{figure}

Figure~\ref{fig:grates_100} shows the growth rates for $Rm = 100$. The cases of $B_0 = 0.001$ and $B=0.01$ are essentially unchanged from that for a purely kinematic dynamo (i.e.\ with $B_0=0$); as $k$ is increased, modes of differing symmetries become dominant (see Courvoisier 2008), leading to a non-monotonic dependence of growth rate on $k$. Increasing the strength of the imposed field has a clear stabilising effect, reducing the growth rates and also decreasing the range of $k$ over which instability occurs. When $B_0=1$ the instability is weak and is confined to a narrow range of $k$ ($0 < k \lesssim 1.2$).

For $Rm = 1000$ and in the absence of an imposed magnetic field, although the most unstable mode is unchanged from that when $Rm=100$ (i.e.\ $k=0.57$ with growth rate $=0.3$), the range in $k$ of unstable modes increases, and the growth rate of unstable modes at higher $k$ increases also. This can be seen by the $B_0=0.001$ curve (essentially equivalent to that when $B_0=0$) in figure~\ref{fig:grates_1000}, which plots the growth rates versus $k$ for $Rm=1000$.
Similarly to the case of $Rm=100$, the effect of increasing $B_0$ is, generally, stabilising, although the picture is a bit more complicated. The mode of maximum growth rate is clearly stabilised, with a slight shift also to smaller $k$ as $B_0$ is increased. The modes at higher $k$ are also shifted with $B_0$, leading to a non-monotonic dependence of growth rate on $B_0$ for certain values of $k$. Inspection of table~3 shows that the linear eigenfunctions are again `magnetically dominated'; the non-monotonicity in $B_0$ of the energy ratio suggests a change in the nature of the most unstable mode for $B_0 \gtrsim 0.1$.

As mentioned above, some care is needed in determining the growth rate for the case of $B_0=0.01$, for which the basic state has a systematic long-time variation. The growth rates shown in figure~\ref{fig:grates_1000} are calculated over a time interval of duration approximately $80$, chosen so as to avoid a disruptive event. During such events, characterised by a rapid change in the basic state magnetic and velocity fields, the linear perturbations do still continue to grow, but at a reduced rate.

\begin{figure}
\vskip 0cm
\hskip 0cm
\epsfxsize 9cm
{\centerline{\epsffile{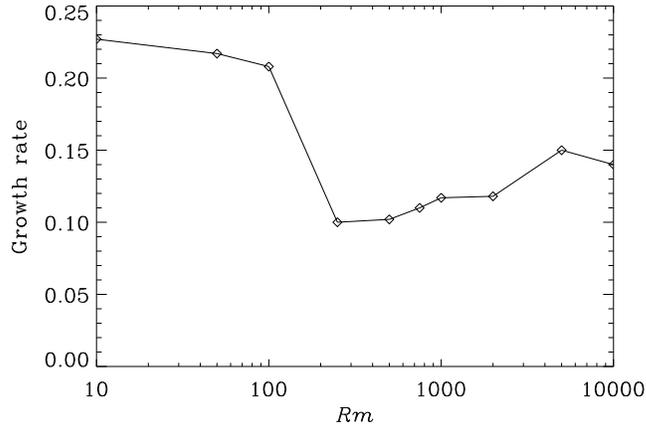}}}
\caption{Growth rate versus $Rm$ for fixed values of $B_0=0.1$ and $k=0.6$.
} 
\label{fig:fast_chk}
\end{figure}

Finally we turn our attention to the problem that motivated the original study of Galloway \& Proctor (1992), namely whether the kinematic dynamo growth rate is bounded away from zero as $Rm \rightarrow \infty$ for a fixed value of the wavenumber $k$ (\textit{fast} dynamo action). Here, in our modified problem, the crucial feature is that at sufficiently high values of $Rm$, \textit{any} imposed field, whatever its strength, will become dynamically significant. Thus the problem in the high $Rm$ limit for any non-zero $B_0$ will necessarily assume a different character from that in the absence of an imposed field. Here we choose to fix $B_0 = 0.1$, a weak field compared with the equipartition value, to set $k=0.6$, essentially the mode of maximum growth rate for $B_0=0$, and to increase 
$Rm$. Figure~\ref{fig:fast_chk} plots the growth rate versus $Rm$. For $Rm \lesssim 100$ the dynamical significance of the magnetic field is weak and the unstable mode is therefore a slight modification to that of the kinematic case. For $Rm \gtrsim 100$ the Lorentz force becomes influential in determining the basic state and the most unstable mode assumes a different character. For $Rm \gtrsim 200$ there is only a slight variation in growth rate with $Rm$; on this numerical evidence we may therefore tentatively conclude that this modified dynamo action is also `fast'. For a smaller value of the imposed field we would expect a similar picture, but with the transition between the two modes occurring at a higher $Rm$.

\section{Conclusion}\label{sec:conc}
We have extended the classical kinematic dynamo problem, in which the velocity is prescribed and one examines the possible exponential growth of a weak magnetic field, in order to consider the linear stability of fully nonlinear MHD states, obtained from a specified $z$-independent forcing and an imposed, initially uniform, horizontal magnetic field $B_0 \bfxhat$. This is feasible since the perturbations, which are fully three-dimensional, can be distinguished from the basic states, which, by construction, depend spatially only on $x$ and $y$. Mathematically the problem involves the solution of linear equations for the perturbations of both the magnetic and velocity fields, in concert with solutions of the two-dimensional nonlinear equations for the underlying basic states.

In the absence of an imposed field the problem reduces to the well-studied kinematic dynamo problem for the Galloway-Proctor flow. The basic state flows formed by the forcing (\ref{eq:force}) acting on an imposed uniform field increase in complexity as $Rm$ is increased, as can be seen from a comparison of 
figures~\ref{fig:energies_100} and \ref{fig:energies_1000}. The stability of these nonlinear MHD states is enhanced as $B_0$ increases, as shown by figures~\ref{fig:grates_100} and \ref{fig:grates_1000}. As is well-known for turbulent flows at high $Rm$, even a weak large-scale field (of equipartition strength divided by some power of $Rm$) can lead to dynamically significant small-scale fields. Such behaviour can be identified in our model, where it can be seen that the imposed field has a dynamical influence on the basic state once $Rm B_0^2$ is $O(1)$. It is however noteworthy that dynamo action can continue in the presence of a significant background field; there is no analogue of the strong quenching of mean field transport coefficients by weak large-scale fields. This is because our model does not represent a two scale process. The normal dynamo properties of the Galloway-Proctor flow cannot be described in terms of an $\alpha$-effect except when $k \ll 1$, and such modes are not preferentially excited. The same remarks hold here, as may be seen from figures~\ref{fig:grates_100} and \ref{fig:grates_1000}. For $O(1)$ values of $k$ the system is really a `small-scale' dynamo, with no distinction between large and small scales.

The motivation for the original study by Galloway \& Proctor (1992) was to investigate whether kinematic dynamo action is fast. In our system, the limit of $Rm \rightarrow \infty$ with $B_0 \ne 0$ is guaranteed to be distinct from the limit of $Rm \rightarrow \infty$ with $B_0=0$; as discussed earlier, an imposed field, however weak, will become dynamically important for sufficiently large $Rm$. (Different effects will occur, however in a real two-scale dynamo.) This is shown in figure~\ref{fig:fast_chk}, in which a change in behaviour as $Rm$ is increased is clearly seen. That said, it does appear from the numerical evidence available, that the instability of the nonlinear MHD state can also be designated as `fast'.

Finally it is of interest to discuss briefly the differences between the approach adopted here, in which we solve the coupled equations (\ref{eq:pert1}) -- (\ref{eq:pert4}), and that considered recently by Cattaneo \& Tobias (2008) and 
Tilgner \& Brandenburg (2008), who considered how a kinematic magnetic field might evolve under the influence of a statistically steady velocity field modified by the Lorentz force. For our system this would be equivalent to solving the single linear equation
\begin{equation}
\frac{\partial \bfb}{\partial t} + \bfU \cdot \nabla \bfb = \bfb \cdot \nabla \bfU + Rm^{-1} \nabla^2 \bfb ,
\label{eq:newpert}
\end{equation}
in conjunction with solving equations (\ref{eq:bs1}) -- (\ref{eq:bs4}) to determine the velocity $\bfU$. For comparison, we have carried out this procedure, which of course is not the same as that adopted in this paper, and find, as do Cattaneo \& Tobias and Tilgner \& Brandenburg, that instability is enhanced. In our model this amounts to modes that are unstable under the full perturbation system (\ref{eq:pert1}) -- (\ref{eq:pert4}) having a greater growth rate when governed only by (\ref{eq:newpert}), and indeed also to some modes that are stable under the full system being unstable in the abridged formulation.

\begin{acknowledgements}
We thank the referees for their helpful comments, which led to the improvement of the paper. We are grateful for the hospitality of the Isaac Newton Institute for Mathematical Sciences, where part of the work for this paper was performed during the programme on \textit{Magnetohydrodynamics of Stellar Interiors}, and also for the support of the STFC. DWH is also grateful for the support of a Royal Society Leverhulme Senior Fellowship, during the tenure of which this work was completed.
\end{acknowledgements}

\end{document}